# Detecting Flaring Structures in Sagittarius A* with (Sub)Millimeter VLBI

*Vincent L. Fish*[1], *Sheperd S. Doeleman*[2], *Avery E. Broderick*[3], *Abraham Loeb*[4], *Alan E. E. Rogers*[5]

[1] Massachusetts Institute of Technology, Haystack Observatory, Route 40, Westford, MA 01886 USA, vfish@haystack.mit.edu

[2] Massachusetts Institute of Technology, Haystack Observatory, dole@haystack.mit.edu

[3] Canadian Institute for Theoretical Astrophysics, University of Toronto, 60 St. George St., Toronto, ON M5S 3H8 CANADA, aeb@cita.utoronto.ca

[4] Institute for Theory and Computation, Harvard University, Center for Astrophysics, 60 Garden St., Cambridge, MA 02138 USA, aloeb@cfa.harvard.edu

[5] Massachusetts Institute of Technology, Haystack Observatory, aeer@haystack.mit.edu

## Abstract

Multiwavelength monitoring observations of Sagittarius A* exhibit variability on timescales of minutes to hours, indicating emission regions localized near the event horizon. (Sub)Millimeter-wavelength VLBI is uniquely suited to probe the environment of the assumed black hole on these scales. We consider a range of orbiting hot-spot and accretion-disk models and find that periodicity in Sgr A* flares is detectable using closure quantities. Our methods are applicable to any model producing source structure changes near the black hole, including jets and magnetohydrodynamic disk instabilities, and suggest that (sub)millimeter VLBI will play a prominent role in investigating Sgr A* near the event horizon.

## 1. Introduction

The compact source Sagittarius A* is believed to host a massive black hole. Observations from radio through X-ray wavelengths find variability on time scales from minutes to hours [1-4]. VLBI imaging at 3.5 mm indicates that the intrinsic size of Sgr A* is approximately 12 times the Schwarzschild radius ($r_{Sch}$) for a 4 x $10^6$ $M_{sun}$ black hole [5]. This size suggests that the emission may be coming from the inner region of the accretion disk. Various models, including orbiting hotspots and jets, have been put forth to explain the variability. The size scale of emission is well-matched to the resolution provided by (sub)millimeter VLBI. Early studies are needed to show the abilities and limitations of (sub)millimeter VLBI and prioritize resource allocation for investigating Sgr A*.

In this work, we investigate non-imaging signatures of Sgr A* from (sub)millimeter VLBI arrays for a variety of disk and hotspot models. A non-imaging approach to analyzing (sub)millimeter VLBI data will be required for several reasons. First, flaring indicates that the source structure changes on timescales short compared to the rotation of the Earth, violating the assumptions of Earth rotation aperture synthesis. Second, at (sub)millimeter wavelengths it is not possible to use the standard phase-referencing techniques regularly employed at centimeter wavelengths. Since individual visibilities will have a fairly low signal-to-noise ratio (SNR) and initial observations may be taken with a 3- or 4-element array, closure relations may not adequately constrain antenna phases for self-calibration. Third, existing

(sub)millimeter arrays provide very sparse coverage of the *(u,v)* plane, so even calibrated data would provide poor image fidelity.

## 2. Models and Methods

Hotspot and accretion disk models including general relativistic propagation effects are described in detail in [6-7]. We consider a suite of models consisting of a single persistent orbiting hotspot, which produces a periodic flare in integrated flux, embedded in a disk around a massive black hole. Models are characterized by black hole spin, hotspot orbital period, disk inclination, and projected major axis orientation and are produced at 230 and 345 GHz. Interstellar scattering is included as given in [8].

We simulate data from an array consisting of Hawaii (James Clerk Maxwell Telescope, Caltech Submillimeter Observatory, and 6 Submillimeter Array antennas phased together), the Heinrich Hertz Telescope at the Submillimeter Telescope Observatory (SMTO), a phased array of eight Combined Array for Research in Millimeterwave Astonomy (CARMA) antennas, the Large Millimeter Telescope (LMT), either a single 12-m-class telescope or a phased array of 10 Atacama Large Millimeter Array telescopes (Chile-1 and Chile-10), the 30 m Pico Veleta (PV) telescope, and the six Plateau de Bure (PdB) dishes phased together. This array, or a subset thereof, is reasonable given current plans for expansion of VLBI into the (sub)millimeter regime. Baselines in the array provide fringe spacings ranging from 20 to 1000 μas; for comparison, $r_{Sch}$ for a 4 x $10^6 M_{sun}$ black hole at 8.0 kpc [9] is 10 μas. Typical system equivalent flux densities are assumed, and we consider all telescopes at 230 and 345 GHz even if capability at both bands is not currently planned. We have assumed a tropospheric coherence time of 10 s [10] and total data recording rates of 1 to 16 Gbit $s^{-1}$ in steps of two.

Closure phases and amplitudes are produced from triangles and quadrangles of baseline visibilities, respectively. Station-based complex gain and clock errors cancel out of the closure quantities, making them robust observables even when array calibration is not possible. The closure phase of a symmetric source distribution is always zero or 180°. Provided that the SNR in a coherent integration time ≳ 1, successive segments of data can be integrated coherently to improve the SNR as $t^{1/2}$, provided that the source structure is not rapidly changing. In the case of unequal baseline SNRs, the SNR of the closure phase is dominated by that of the weakest baseline. Further details can be found in [11].

## 3. Results

Present and future (sub)millimeter VLBI arrays will be able to detect the signatures of changing structures on angular scales of tens to hundreds of microarcseconds. Figure 1 shows simulated 230 GHz data as would be detected over a two-hour flare. Varying deviations from zero closure phase, indicative of changing asymmetric structure, are evident on most or all triangles, depending on the model. Small triangles, such as Hawaii--SMTO--CARMA, may not have sufficient angular resolution to detect changes in the source structure. Larger triangles, such as those to Chile, will resolve out much of the source flux and may require the use of Chile-10 to produce a convincing detection. Simulated data at 345 GHz are similar.

If the source of flaring events is a hotspot in a stable circular orbit, the period of the orbit can be easily extracted from the autocorrelation of the closure phase time series, even if the source is only marginally detected in each coherent integration. As seen in Figure 1, closure phases will not repeat exactly because the rotation of the Earth will change projected baseline lengths and orientations, but the variation in closure phase from orbit to orbit will not be large provided that the orbital period is on the order of tens of minutes or less. On the other hand, even if the mechanism that produces flares in Sgr A* is aperiodic, it will produce asymmetries in the source structure that should be detectable via closure phases on one or more of the triangles available for (sub)millimeter VLBI.

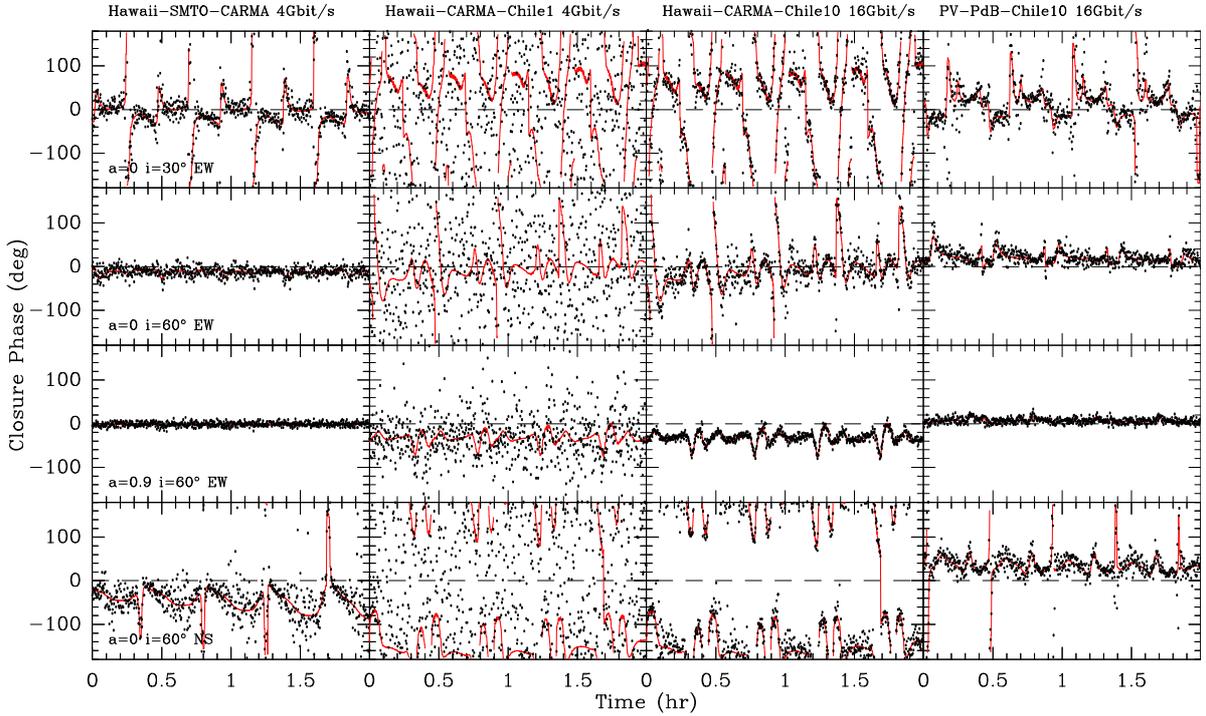

Figure 1: Simulated data at 230 GHz (points), with noiseless models in red. Models are denoted by black hole spin ($a = 0$ or 0.9 times maximal), inclination of spin axis to line of sight, and projected major axis orientation. The hotspot orbits at the innermost stable circular orbit for $a = 0$ and a comparable radius at $a = 0.9$. Two hours of data, corresponding to 4.5 periods, are shown. The second and third columns illustrate why Chile-10 may be important.

## 4. Conclusions

Now that Sgr A* has been detected on VLBI scales at 230 GHz [12], it is clear that future efforts should be directed toward improving the sensitivity of potential (sub)millimeter VLBI systems. In the short term, it will be important to focus efforts on producing phased array processors to leverage the existing collecting area on Mauna Kea and at the CARMA site. Ongoing development of ultrahigh bandwidth digital backend systems will provide aggregate bit rates of 4 Gbit/s presently and 16 Gbit/s within the next few years. Since there is an SNR threshold below which closure phases do not integrate coherently, burst mode recording capability may be important. Cryogenic sapphire oscillators may provide greater frequency stability than is currently available with hydrogen masers, which will be especially important for investigations at 345 GHz and higher. It will also be critical to observe simultaneously with four or five telescopes. The number of independent closure quantities (phases and amplitudes) grows quickly with additional telescopes, and the additional data may be critical for confirming marginally-detected periodicity.

It is important to note that these methods and results are applicable to any model of changing source structure on scales near the event horizon in Sgr A*, whether that be due to a hotspot in a stable or spiraling orbit, a jet, magnetohydrodynamic disk instabilities, or some other mechanism. Angular scales probed by VLBI at 230 and 345 GHz range from 2 to 100 $r_{Sch}$. All reasonable models of Sgr A* produce inherent asymmetries on these scales in addition to the asymmetries that are inevitably produced by strong gravitational lensing. With planned instrument upgrades and technological developments currently underway, VLBI at (sub)millimeter wavelengths will very soon be an invaluable tool for exploring the region around the event horizon in Sgr A*.

## 5. Acknowledgments

The high-frequency VLBI program at Haystack Observatory is funded through a grant from the National Science Foundation.